# The Argument for Meta-Modeling-Based Approaches to Hardware Generation Languages

Johannes Schreiner*, Daniel Gerl*†, Robert Kunzelmann*†, Paritosh Kumar Sinha*, Wolfgang Ecker*†
*Infineon Technologies AG, Neubiberg, Germany
‡Technical University of Munich, Munich, Germany
Email: firstname.lastname@infineon.com

*Abstract*—The rapid evolution of Integrated Circuit (IC) development necessitates innovative methodologies such as code generation to manage complexity and increase productivity. Using the right methodology for generator development to maximize the capability and, most notably, the feasibility of generators is a crucial part of this work. Meta-Modeling-based approaches drawing on the principles of Model Driven Architecture (MDA) are a promising methodology for generator development.

The goal of this paper is to show why such an MDA-based approach can provide extremely powerful generators with minimal implementation effort and to demonstrate that this approach is a superior alternative to the most advanced hardware generation languages such as SpinalHDL and Chisel. For this purpose, this paper provides an in-depth comparison of the Meta-Modeling approach against these hardware generation languages, highlighting the unique advantages of a Meta-Modeling-based approach and summarizes the benefits.

*Index Terms*—Hardware Description Languages (HDLs), Hardware Generation Languages (HGLs), Meta-Modeling, Model Driven Architecture (MDA),

## I. INTRODUCTION

Continuous productivity increases in hardware design are essential to keep pace with the continuously growing complexity in semiconductors. More fundamentally, productivity increases are essential for our health, comfort, well-being, and all the other good things the humankind achieved over the last 100 years. Unfortunately, productivity is stagnating world-wide, with productivity increases down from an 8% high in the 1950s to around 1% today. This is a disaster since more productivity is needed to solve the existing and also new challenges. These include climate change, more rapidly and frequently evolving diseases, and 735 million malnourished people which equals about 9% of the world's population.

In the field of semiconductor design, a solution to the productivity disaster will not only enable more products to be designed in shorter time and with less effort but also to make chips at affordable prices that make our lives easier, safer and greener.

This paper discusses a promising technology known as hardware generation. The concept of employing generation as an innovative approach to chip design has been proposed by Nicolic [1] and Shacham et al. [2]. This approach challenges conventional methods by advocating for the codification of the entire body of the designer's knowledge that culminates in the creation of a single chip, instead of merely codifying the chip itself.

The ramifications of this shift are significant, opening up substantial opportunities for generator-level reuse. By leveraging the capabilities of these chip generators, a multitude of diverse chips can be produced without the necessity for additional manual design work. Each of these generated chips can be optimized for distinct applications and different trade-off criteria, thereby, making the approach highly adaptable and flexible.

Facilitating the transition from the development of static models to generators requires a range of activities aimed at increasing efficiency. These encompass the implementation of strategies and techniques to streamline the development process, promote effective knowledge transfer, and capitalize on the inherent benefits of the generative approach. The objective is to create a robust, dynamic, and scalable system of chip generation that not only meets the demands of the present but is also capable of adapting to the evolving needs of the future.

However, while the benefits of chip generation are clear, there are challenges to the adoption of this technology. One of the primary obstacles is the 'Generator Gap', a problem which we will explore in the subsequent section.

## II. PROBLEM: THE GENERATOR GAP

In the realm of chip design, the application of generators has been explored as a promising strategy for enhancing productivity and managing complexity. However, the effective implementation of generators poses a challenge, as depicted in Figure 1. The ideal scenario is to have highly abstract input data or specifications for the generator, while the generated artifacts need to be at a low level of abstraction, suitable for output tools and workflows.

As the aspiration to make generators more ubiquitous and sophisticated intensifies, the levels of abstraction on the input side rise, with an ideal target of full-chip generation from high-level specifications. Conversely, the reduction in the level of abstraction in the target views is necessitated by the requirement to generate views that factor in low-level physical properties of the targeted hardware, such as timing budgets or low-power mechanisms. This results in an escalating complexity that generators need to bridge, which makes the development of generators challenging and costly.

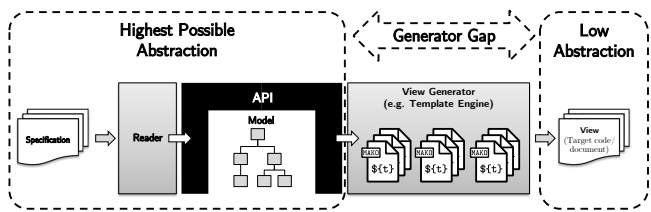

Figure 1. The Generator Gap in Simple Generators

This paper coins the term 'Generator Gap' to refer to the increasing divide between specifications and low-level target views. The main challenge in bridging this gap lies in the complexity of the generator development. Powerful code generation approaches should ideally bridge a large Generator Gap, providing the ability to generate target views of a low level of abstraction from specification inputs at the highest possible level of abstraction. Identifying novel ways to make the development of code generators easier is crucial to increase the level of abstraction for inputs of existing generators and to discover new areas where model-centric automation can be applied.

The complexity of bridging the Generator Gap stems from the following sources:

1) The intertwining of business logic, used to analyze, understand, and automatically refine the input model (i.e. specification) to the level of the target view, with the logic that emits the language or format of the target view. The complexity escalates particularly when the language of the generated target view is similar or identical to that of the generator.
2) The need for configurable views generation leads to increased complexity, as it has to be considered alongside the business logic responsible for automatically refining the specification. In many scenarios, the number of configuration flags required for an individual generator significantly impacts the code size and complexity of the generator.
3) The close intertwining of generators with the target code they were designed for introduces two main issues: First, the duplication of business logic when different kinds or styles of target code shall be generated. This increases the overall complexity of the generator development task. And second the challenge of adding new generators or modifying existing ones for a new target platform.
4) The close intertwining of generators with the target domain: Besides Register-Transfer Level (RTL) code also other collaterals such as firmware, testbenches, virtual prototypes, formal properties and documentation are needed and contribute to the overall design. Only putting one of these collaterals in the generator chain in focus substantially reduces the potential benefit that can be gained from generation.

The complexity and issues outlined above present significant barriers to progress and prohibit the expansion of the use of model-centric code generation across the software and hardware development landscape. Addressing these challenges is vital to unlock the full potential of chip generations as detailed in the previous section. A solution to these issues requires a systematic, structured, and reusable approach. Sections III and IV introduce design language inherent features enabling generic design, preprocessors as well as script-based approaches and Hardware Generation Languages (HGLs). Section V finally introduces the Model Driven Architecture (MDA) and its potential to address these complexities and to pave the way for further automation.

## III. Traditional Generation Approaches and Their Limitations

This section first introduces the generation and configurable features available in Hardware Description Languages (HDLs) that are currently used by the industry and the established standard for digital design. It then introduces the concept of pre-processors, Embedded Domain-Specific Languages (EDSLs)[1], which is an approach to the development of custom languages that is heavily applied to develop custom, language-based code generation methods. Eventually, this section will present two important stages of evolution in the development of generator languages as EDSLs:

First, it shows how the EDSL concept has been applied to significantly extend the configuration possibility of existing industry standard HDLs with their underlying simulation semantics.

Second, it prepares the introduction of a family of so-called Hardware Construction Languages (HCLs), also referred to as Hardware Generation Languages (HGLs). Those are a group of EDSLs that are developed from the ground up with a focus on hardware generation and the need for productivity during generator development. Section IV introduces and discusses those in more detail.

### A. Configurability in VHDL, Verilog, and SystemVerilog Designs

This section discusses the capacity of standard HDLs namely VHDL, Verilog, and SystemVerilog to create flexible, configurable designs. It explores the industry-standard approaches and their limitations.

*1) Preprocessors:* Verilog and SystemVerilog offer preprocessing capabilities, defined as compiler directives in the IEEE 1364-1995 Verilog standard, and expanded upon in subsequent Verilog standards and the combined IEEE 1800 standard for Verilog and SystemVerilog [3, 4]. The preprocessing functionalities in these languages are similar but more restricted compared to their counterparts in C and C++ programming languages [5].

---

[1] A non-embedded DSL—or simply DSL—defines a complete new language with new syntax, grammar, and semantic. While at the start slightly easier to use it requires a higher effort to build an environment for the language and to get used to it.

Preprocessing directives are frequently used in the industry to manage different representations for design and synthesis of Verilog or SystemVerilog models. Typically, this is achieved with simple `` `ifdef `` or `` `ifndef `` blocks and additional definitions introduced through the tool or compiler interface.

Another common use is 'include guards', where an entire file's content is wrapped in an `` `ifndef `` statement that ensures the content will be included only once, regardless of the frequency of the `` `include `` directive encounters.

Despite these functionalities, the preprocessing capability of HDLs is not intended for hardware generation. This is underscored by the absence of an `` `if `` in SystemVerilog directives [5]. Similarly, VHDL, though supporting conditional compilation directive in its latest standard, does not have vendor support for this extension [6]. Hence, current HDLs are not seen as tools ideally suited to create highly adaptable, reusable hardware generators.

*2) Parametrization and Generation Constructs:* Both VHDL and SystemVerilog support parametrization, referred to as 'Parameters' in SystemVerilog and 'Generic' in VHDL. These parameters are used to control module properties and can be of various types, such as Boolean flags, string identifiers, or integer values defining the size of internal elements.

In conjunction with parameters, the `generate` construct in both SystemVerilog and VHDL enables configurability of design internals. This provides support for loops, `case` statements, and simple `if elseif else` blocks, all of which can be nested within a generate block.

For example, in SystemVerilog, a module can use a `generate` block to support a configurable number of channels. The number of submodules instantiated will vary based on the number of supported channels, which can be mapped to individual bits or subarrays of the inputs using a loop variable.

*3) Limitations:* The capabilities of VHDL and SystemVerilog, while enabling the creation of configurable HDL modules to a degree, encounter significant limitations. One primary constraint is the inability of the `generate` feature in both languages to introspect and analyze the part of the design it is embedded in. Consequently, the scope of information that can be utilized within the `generate` context is restricted to the parameters and their interrelationships.

Moreover, `generate` blocks must be part of the module body, preventing the conditional addition or removal of ports, or the generation of a flexible number of module ports. Furthermore, the constraint of accepting only simple types as parameters limits the complexity of input parameters that can be used in Verilog and VHDL.

Preprocessors in these languages as already introduced and discussed before, despite their potential, lead to unwieldy syntax and challenging readability when used for building generators. The support for complex constructs is notably sparse, with configuration being passed in through primitive parameters, either combined into files containing `define` statements or individually through the interfaces of Electronic Design Automation (EDA) tools interpreting the code. This is fundamentally less capable than even basic script-based generation methods.

The integrated parametrization and generation constructs in these HDL languages provide a more structured approach to creating generic and configurable HDL designs, albeit with certain limitations. They lack support for any design introspection and have clear limits to their generator capabilities. The strength of such an approach lies in its integration into the HDL language standards and widespread tool support in the EDA ecosystem.

However, it is evident that both VHDL and SystemVerilog were designed with the primary purpose of describing single design instances, with configurability being an afterthought. Such limitation in flexibility and configurability, in contrast to the capabilities of the MDA approach presented later in this paper, indicates the inability of traditional HDLs to meet more complex configurability needs.

*B. Advanced Preprocessing and Script-Based Generation*

Preprocessing tactics and alternative language descriptions of HDL code have been devised to combat the restricted functionality of mainstream HDLs employed within the industry. The most elementary method to achieve this was previously explained in Section III-A. This approach employs scripting languages, such as Perl and Python, for preprocessing tasks [7, 8]. This technique offers a substantial improvement over the preprocessors specifically designed for HDLs: These scripting languages are widely comprehended and supported, enhancing readability for an extensive demographic of developers.

In more sophisticated methods, template engines are incorporated, enabling developers to work primarily in their preferred HDL, and only interject *preprocessor directives* if necessary. This morphs into a substantial upgrade in readability in contrast to scripts filled with numerous print statements that create static elements of the target view. Genesis2 illustrates this approach effectively [9]. This code generation tool employs Perl and a Perl template engine to integrate Perl preprocessing logic within Verilog files.

Special `*.vp` files can carry Verilog code which is blended with Perl code. Listing 1 shares an example of such file from the original authorship of Genesis2. The Verilog module is punctuated with Perl commands and control statements. Subsequently, the framework executes and morphs the Perl code, and then produces `*.v` Verilog files from it. To illustrate, Lines 15-18 contain a Perl `for` loop that will repeat the content of Lines 16 and 17 multiple times, dependent on the value stored in `$N`. For each of the repeated lines, Perl expressions enclosed in `'` are evaluated and replaced with their evaluation result [9].

This outlined template engine shares similarities with the Meta-Modeling-based generation flow pictured in Figure 1. However, there is a notable difference: The generation process mentioned here does not apply to the advanced Meta-Models and the Meta-Model-based generation which is typical for cutting-edge Meta-Modeling-based generation flows.

```
1  //; use POSIX ();
2  module 'mname()'
3   (input logic ['$N-1':0] pp['$N-1':0],
4    output logic ['2*$N-1':0] sum,
5    output logic ['2*$N-1':0] carry
6   );
7   //; my $height = $N;
8   //; my $width = 2*$N;
9   //; my $step = 0;
10  // Shift weights and make pps rectangular
11  logic ['2*$N-1':0] pp0_step'$step';
12  assign pp0_step'$step' =
13   {{('$N'){1'b0}}, pp[0]};
14
15  //; for (my $i=1; $i<$N; $i++) {
16  logic ['2*$N-1':0] pp'$i'_step'$step';
17  assign pp'$i'_step'$step' =
18   {{('$N-$i'){1'b0}},pp['$i'],{'$i'{1'b0}}};
19  //; }
20  ...
21  assign sum = pp0_step'$step'['2*$N-1':0];
22  assign carry = pp1_step'$step'['2*$N-1':0];
23 endmodule : 'mname()'
```

Listing 1: Sample Script-based Preprocessor from the Primary Research that Developed Genesis2 [9]

This technique inherits a limitation from the Meta-Modeling-based generation flow and from the development of adaptable designs which rely solely on the features of VHDL and SystemVerilog – the incapacity to introspect into the generated design during the generation flow. Both methodologies generate the design in the target HDL, which is not employed during the generation flow but only upon its conclusion. Furthermore, these approaches are hampered by the issues identified as the Generator Gap, which appear even more pronounced due to the absence of the structure provided by Meta-Modeling within the generation flow. However, it is important to state that preprocessor approaches allow more genericity than HDLs and have a quite flat learning curve as code that should be made generic is directly visible.

## IV. Exploration of Hardware Generation Languages and Their Constraints

In this section, we delve into the realm of HGLs. It is noteworthy that this class of languages is often alluded to as Hardware Construction Languages (HCLs) in related literature. Furthermore, certain publications and languages employ the term Hardware Description Languages (HDLs) when referring to these methodologies. The term HDL is indicative of the fact that these languages could potentially replace conventional HDLs such as VHDL and SystemVerilog in the development of synthesizable hardware descriptions. However, it overlooks a crucial aspect: The primary objective of these languages is to articulate generators for diverse hardware configurations rather than describing individual instances. To stress the focus these languages place on creating flexible generators for synthesizable designs, this paper consistently refers to these languages as HGLs.

The research conducted in the sphere of HGLs hinges strongly on the concept of constructing generic design generators, as opposed to delivering single instances of a design. These languages are the logical successors to those discussed in the preceding section, although the demarcation between HDLs and HGLs implemented as EDSLs is rather blurred.

In this study, we adopt the following as defining characteristics of EDSL-based HDLs: An HDL based on EDSLs is a language primarily designed to extend or provide capabilities for configurability, parametrization, and generation of an existing HDL. The widely used HDLs VHDL, Verilog, and SystemVerilog, which receive substantial support from the EDA industry, were originally designed and utilized exclusively for specification as well as documentation (VHDL) and simulation (Verilog) purposes. The semantics they employ to describe hardware is rooted in event-driven simulation. These languages were later adapted for synthesis purposes. While design engineers adopt an *implementation thinking* approach, focusing on hardware semantics, they need to transpose this mental model onto the event-driven simulation semantics of HDLs. Synthesis tools must then deduce the design intent from these descriptions. EDSL-based HDLs leverage the same underlying simulation semantics and event-driven paradigm as their parent HDLs, and consequently, encounter the same predicament.

For EDSL-based HGLs, this core characteristic is not applicable. In this paper, we define HGLs as languages devoid of any influence from the simulation semantics of VHDL and Verilog. Consequently, the following two characteristics are central to these languages:

1) They employ the semantics of hardware and the design-centered thought process as their foundational semantic model.
2) They are designed for generation and meta-hardware description. HGL code does not serve as a model of the hardware but as an executable program that, when executed, generates a description of the hardware.

### A. Examining Chisel and SpinalHDL

Today, in the domain of HGLs, Chisel and SpinalHDL stand out as the most prominent representatives. Chisel was originally conceived as a part of a UC Berkeley research initiative with the objective of enabling hardware designers to script high-level, synthesizable hardware descriptions [10].

Chisel is an EDSL that leans on the Scala programming language, a versatile language with robust support for functional and object-oriented programming, compiled to run on the Java Virtual Machine (JVM). Employing Scala as the host language allows Chisel to leverage the robust interoperability with libraries and compatibility with tools within the Java ecosystem. Another advantage of Scala is its robust support for functional and object-oriented programming, making it an ideal foundation for devising an Embedded DSL. In fact, the Scala language was deliberately designed with the EDSL use-case in mind [10].

The inaugural version of the Chisel language was unveiled in 2012. The accompanying research established that Chisel is suited for constructing complex and flexible hardware generators, offering superior efficiency and a higher level of abstraction than SystemVerilog. It also demonstrated that Chisel could generate RTL designs that were competitive in terms of their area and power usage [10].

The prospective productivity benefits of implementing Chisel also sparked interest for industrial applications. However, the initial Chisel version lacked vital features required for cutting-edge industrial designs.

Noteworthy inadequacies of the initial Chisel version included the absence of multiple clock domain support within one module, no support for falling edge clocks and active low resets, lack of clock enable support, and the absence of support for blackbox modules for incorporating legacy circuits described in standard HDLs. Moreover, several shortcomings in convenient reset handling and syntactical inconveniences limited the language's benefits. A crucial flaw of the original Chisel version was its failure to strictly enforce hardware semantics: `when` blocks used by Chisel to describe multiplexer structures did not ensure that all values were always described. Thus, it was possible to generate RTL that contained asynchronous signals that were not assigned in every way. This behavior essentially introduced event-driven simulation semantics into the language. The inability of the Chisel project to address these issues in a timely manner eventually resulted in the creation of SpinalHDL [11].

After the SpinalHDL fork from Chisel, both languages have been developed in parallel and are employed in academia and, to some extent, in industry. It is essential to note that all of the shortcomings identified by the SpinalHDL project in 2016 have been resolved and no longer apply to the latest version of Chisel3. Over time, the differences between the syntax, feature set, and library of SpinalHDL and Chisel have expanded. Within the scope of this paper, only Chisel will be introduced as it is the more popular language and practical for use in both academic and industrial contexts and is the subject of more ongoing research and development. The following elaborates on the internal semantics of Chisel and the architecture and infrastructure of the backend.

### B. The Semantics of Chisel

Semantically, Chisel is a design-centric HGL. It is by construction free from any event-driven simulation semantics. This means that every line of code written in Chisel either describes the design or contains generator logic responsible for configuring and customizing the generated design.

When Chisel code is executed, a hardware design description is built. This construction of the design has several important characteristics:

- Due to the absence of any simulation semantics, all stateful elements such as Registers and Memories always need to be explicitly instantiated and connected and are never inferred. Generator code that would lead to the incomplete definition of multiplexers due to missing assignments of combinatorial logic values will not run in Chisel, i.e. no hardware description for simulation or synthesis is generated.
- Connectivity is described using wires and connections. There are no signals or signal semantics, just wires in the constructed hardware model. Consequently, there cannot be any *assignment* of values. The `:=` operator, which is sometimes referred to as the assign operator in the Chisel documentation and publications is actually a pure connection and connection modification operator.
- Combinational logic is defined using wires and expressions. In this context, every wire and port (their generalization in the Chisel core is referred to as a node) must have a driver connected to it. Thus, undefined states do not exist.

The description of combinational logic in Chisel is quite intuitive and will not be further detailed here. An essential aspect of it, the creation of multiplexers is, however, intertwined with the connectivity creation of the Chisel language. Both aspects will be described in the following.

As mentioned before, Chisel supports ports, literals, and wires and generalizes these elements as *nodes*. Connections between these nodes are created using the `:=` connection operator. The example in Listing 2 contains three nodes: `io.input`, `io.output` and `myNode`. Two out of those nodes need to have a driver inside the module (`io.output` and `myNode`). The example contains two noteworthy aspects: First, the two connections created in Lines 11 and 14 seem contradictory as they would connect two drivers to the same node. This is resolved by the definition of the `:=` operator in Chisel. This operator is defined to do two things: First, it will also disconnect any other driver connected to the left-hand side node if there are any connections already. Second, it will connect its right-hand side driver to its left-hand side node instead. In the given hardware model, `io.output` will therefore be connected to `myNode` which in turn will be only connected to `io.input`, not to the literal.

```
class SampleModule extends Module {
  val io = IO(new Bundle {
    val input = Input(UInt(8.W))
    val output = Output(UInt(8.W))
  })

  val myNode = Wire(UInt(8.W))
  io.output := myNode

  // connect a constant to the node myNode
  myNode := 255.U
  // disconnect driver 255.U from myNode
  // and connect io.input to the node MyNode
  myNode := io.input
}
```

Listing 2: Sample of a Chisel Module with Conflicting Connection Statements

This behavior is illustrated by Figure 2, which shows the internal connectivity model the Chisel library builds from

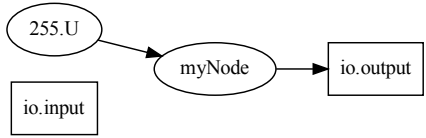

(a) Model after Execution of Line 11

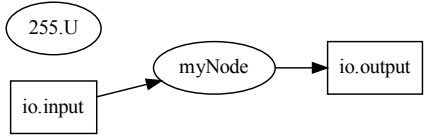

(b) Model after Execution of Line 14

Figure 2. Hardware Models Generated by Listing 2

```
1  class SampleModule extends Module {
2    val io = IO(new Bundle {
3      val input = Input(UInt(8.W))
4      val output = Output(UInt(8.W))
5    })
6
7    val myNode = Wire(UInt(8.W))
8    io.output := myNode
9
10   myNode := 255.U
11
12   when(io.input < 128.U) {
13     myNode := io.input
14   }
15 }
```

Listing 3: Sample of a Chisel Module with Multiplexer Connection

the executed code. Sub-Figure (a) shows the status of the internal model after the first connection in Line 11 has been executed. Sub-Figure (b) then shows how the connectivity model changes when Line 14 has been processed. What happens is that the connection from Line 11 is removed again, resulting in a model that is functionally equivalent to a model where the line of code is commented out.

To fully understand the semantics of Chisel, it is also important to understand the impact of the conditionals `when`, `elsewhen`, and `otherwise`. The Chisel `if` constructs behave similarly to the `if` constructs in SystemVerilogs `generate` blocks: They are evaluated at runtime of the generator and will statically decide whether a certain part of the generator will be run or not, i.e. whether a certain piece of hardware will be generated or not. The `when` constructs on the other hand are conditionals that will be included in the design. In other words, these constructs are responsible for generating multiplexers with conditions as their select signals. On the Chisel language level, the `when`, `elsewhen`, and `otherwise` blocks impact the design by modifying the behavior of all the connectivity operators `:=` used inside these blocks. If the connectivity operator is embedded in one or more of these conditional statements, it will no longer simply disconnect any driver from the left-hand side element and connect the right-hand side element. Instead, a connectivity operator embedded inside one or more `when` statements instantiates a multiplexer. In other words, the `when`, `elsewhen`, and `otherwise` blocks are a convenient and flexible way to introduce multiplexers into the connection path.

Listing 3 is a modified version of Listing 2 where the connectivity operator for the connection from `io.input` to `myNode` is embedded in a conditional `when`-block. Inside this `when`-block, the connectivity operator will exhibit the following behavior: First, it disconnects the current driver (`255.U`) from `myNode`. Next, it creates a new multiplexer in the design and connects the output of this multiplexer to `myNode`. Finally, it connects the right-hand side of the connection operator to the true path of the multiplexer and the original driver to the false path of the multiplexer.

In case of nested `when`-blocks, it generates hardware to evaluate the boolean value for the and-combined condition of all wrapping when clauses the connection operator is embedded into (in this example just one clause and one condition). This condition is then connected to the select input of the multiplexer. Figure 3 shows the hardware model generated by this example: Both `io.input` and the literal `255.U` are connected to a multiplexer. The select of the multiplexer is connected to a less-than comparator (`lt`) which compares the input `io.input` to the constant `128` provided in the Chisel code.

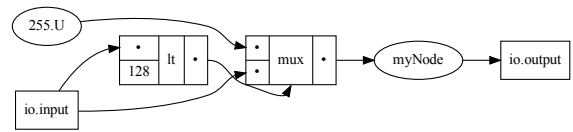

Figure 3. Hardware Model Generated by Listing 3

If the code is modified from Listing 3 and the statement in Line 10 is not hidden in a comment, this will remove the connection between the literal `255.U` and the input of the multiplexer created by the `when` block in combination with the assignment in Line 13. The input node of the multiplexer would therefore not be connected to any driver, violating the Chisel rule that every node needs to have a driver and the underlying hardware semantic. Running this Chisel generator will fail with an exception highlighting the issue and not generating HDL output.

The last important point regarding the semantics of Chisel is its stateful elements. For every register that is created in Chisel, the data output node and the data input node are connected. Following the semantics of the connectivity operator `:=`, an assignment to the register that is not wrapped inside a `when` statement will disconnect the data output from the register and attach the right-hand side of the assignment instead. The implications of this definition and the Chisel semantics described above are illustrated in Listing 4, which is a modified version of the last example in Listing 3.

```
class SampleModule extends Module {
  val io = IO(new Bundle {
    val input = Input(UInt(8.W))
    val output = Output(UInt(8.W))
  })

  val myReg = Reg(UInt(8.W), init: 0)
  // replace the wire with a register
  io.output := myReg

  // myReg := 255.U // comment out this line

  when(io.input < 128.U) {
    myReg := io.input
  }
}
```

Listing 4: Sample of a Chisel Register and Port (Nodes) without Driver

For this example, the wire in Line 7 has been replaced and the connection created in Line 11 has no longer been hidden behind a comment. Because of the replacement of the `Wire` with a `Reg` and because the implicit connection between register output and register input is automatically created for every register, this example generates HDL despite the missing default assignment. The assignment in Line 14 removes the data output of the register from its data input and attach it to the newly created multiplexer instead.

### C. Advantages and Limitations

One of the key advantages of Chisel lies in its relative simplicity compared to the traditional, industry-standard HDLs. A big part of this comes from Chisel's semantic clarity: It is fundamentally a design language based on the underlying hardware semantics and free from event-driven or simulation-based concepts. It does not infer stateful elements and every Chisel statement ultimately creates hardware and connections between this hardware. Moreover, it has a very concise syntax, making it easy to write and read generator code.

Chisel has gained high popularity over the last years, with plenty of research work, open source IP components, learning materials, and a vivid community developing based on it. Recurring conferences and some limited industry adoption also have helped the Chisel development.

Nonetheless, we identify several high-level drawbacks of the Chisel approach:

*1) Common Weakness with HDLs:* The discussions around the `when`-statement also shows that the HGLs are not fully decoupled from weaknesses introduced by HDLs. Conceptually multiplexers shall be inferred but this is neither explicitly stated in the code nor directly supported in the language. Instead checks are still required to restrict the use of the `when`-statement to make sure that this finally may end-up in a multiplexer.

*2) Complex Host Language:* In terms of the initial learning curve, a significant drawback of the Chisel approach becomes visible. Chisel suffers from the relative obscurity of the underlying host language Scala: None of the common rankings published for programming languages places Scala in the top 20 languages [12, 13]. Scala is a complex multi-paradigm language that has a significantly higher entry barrier than other general-purpose languages. Scala is not a language that is already part of science and engineering university curricula or commonly used for scripting and general-purpose programming and automation. When it comes to available libraries, Scala code can easily access the Java ecosystem [14]. A set of libraries as for numerics, data science, and artificial intelligence are particularly useful for automation in the field of Integrated Circuit (IC) design. These libraries are easy to access from native compiled languages and Python, which typically acts as the front-end language for the AI and ML ecosystem. Bindings to the Java ecosystem are only available in a limited manner. In Section III-B, we identified a flat learning curve of script language inlets and availability of libraries benefits of EDSLs. Given the properties of the Scala language, this benefit does not fully apply to Chisel [10].

*3) Lack of Visible and Clearly Defined Underlying Meta-Model:* During the execution of Chisel code, the data structure is built from the Chisel language's instructions. This data structure is then processed and written out as Verilog for synthesis. The easiest path to understanding and conceptualizing Chisel is to understand the language definition. Following this learning path (getting the syntax and semantics of the language instead of learning the structure and semantics of the underlying model) makes the initial learning of Chisel easier, however, comes with a caveat: This learning path does not require an understanding of the model underlying the Chisel-generated data structures.

The *Complex Host Language* of Chisel and its properties, operators, and constructs—which may be unknown to the developer of Chisel code—are closely interwoven with the operators and constructs that the DSL puts on top of it, making it difficult for the user to understand what actually happens when the code is executed. The lack of a visible and clearly defined underlying Meta-Model or more precisely the lack of awareness of this Meta-Model to define the intermediate makes it difficult to understand what the individual Chisel statements do to modify this Meta-Model. Clear visibility on the existence of the Meta-Model and the actions done by the executed generator code to this Meta-Model helps developers to understand that they are writing generators instead of writing HDL code.

We consider two points as key features of EDSL-based HGLs:

1) The ability to perform introspection on the existing data structures describing the design *during* the execution of the generator

2) The ability to apply transformations on the generated data structures

These two features cannot easily be achieved due to the lack of a visible, clearly defined underlying Meta-Model: Chisel was fundamentally designed as a language and not as a tool to define a model. The language specification hides the model underlying these data structures, making the development of introspection or transformation-based approaches more difficult. It is possible to perform introspection on generated structures during the execution of the generator and to apply any transformation to them. This, however, requires an understanding of the internals of the Chisel library and compiler. Such an understanding is not at the center of a language-based approach and is therefore difficult to obtain. Moreover, the data structures this will work on are outside of the specification of Chisel and the developed code may fail in later versions of the language.

We also see the model and the understanding of the Meta-Model as a central point to understanding the semantics of a language. The absence of this Meta-Model in the Chisel learning path makes it difficult to separate the (very nice and clean) semantics of the Chisel language from the semantics of the target view (event-driven simulation language such as SystemVerilog) without a focus on the underlying generated model.

A further disadvantage of Chisel is the relatively low level of the underlying internal model that Chisel data is mapped onto. The section on the semantics of the language already described that Chisel code is basically broken down into a graph of nodes, with registers, combinational logic, and submodule instantiations as its key elements. Chisel notably does not contain common higher-level constructs such as finite state machines that are part of the design thinking of hardware developers. This can be a significant limiting factor for transformations and analysis on the model built by Chisel code.

*4) Assessment of Limitations by the Chisel Team:* Some of these drawbacks are also identified and acknowledged by the authors of Chisel [15]. They correctly note several limitations that all originate from a non-model-based, language-centric approach:

1) "Writing custom circuit transformers requires intimate knowledge about the internals of the Chisel compiler" [15]. We attribute this limitation mainly to the *Complex Host Language*.
2) "Chisel semantics are under-specified and thus impossible to target from other languages" [15]. We attribute this limitation to the *Lack of Underlying Meta-Model*.
3) "Error checking is unprincipled due to under-specified semantics resulting in incomprehensible error messages" [15]. We attribute this limitation to the *Lack of Underlying Meta-Model* with a formal, Meta-Model-based description of legal model states.
4) "Learning a functional programming language (Scala) is difficult for RTL designers with limited programming language experience" [15]. From our point of view, this originates from the *Complex Host Language*.
5) "[C]onceptually separating the embedded Chisel HDL from the host language is difficult for new users" [15]. From our point of view, this originates from both the *Complex Host Language* and the *Lack of Underlying Meta-Model*.

*D. FIRRTL*

The developers of Chisel have addressed the shortcomings they identified with the introduction of the Flexible Intermediate Representation for RTL (FIRRTL). FIRRTL is described by the authors as an HDL that represents the elaborated Chisel circuit, having many syntactical and semantic similarities with the Chisel language, however, do not contain any of the meta-programming capabilities provided by the Scala host language [15]. The FIRRTL code is what is generated from the Chisel internal data structures describing the hardware. From the point of view of this paper, FIRRTL is an intermediate model with a well-defined Meta-Model that describes an elaborated Chisel design. In other words, it is the hardware-centric model built by the execution of a Chisel generator.

FIRRTL does address the shortcomings originating from the *Lack of Underlying Meta-Model* we described in the previous section:

- It provides a clean platform for applying transformations.
- It helps developers to conceptually separate the Chisel language from the host language as they can see what the evaluated circuit looks like without the host language generator constructs, yet still in a syntax similar to that of Chisel.
- It can be directly generated without using the Chisel generator if a different path, absent of Chisel HGL has to be used to generate circuits.

While it can be clearly seen that FIRRTL is removing these shortcomings and making the Chisel flow more similar to the flow presented in this work, there are a few remaining limitations and problems with the approach:

- The FIRRTL model is not the internal model of the Chisel language. Thus, FIRRTL does not support the process of introspection on the Chisel model at runtime of the Chisel generator. The FIRRTL model is similar, yet not identical to the internal model of the Chisel language. It does therefore not significantly ease the process of introspection on the Chisel model right when the Chisel generator is executed.
- Development of generators in Chisel is still the development of generators in a language defined for hardware generation, it is not the development of a transformation or the explicit generation of a target data structure. This target data structure is then transformed into FIRRTL. However, the developer is not necessarily aware of this first layer of data structure that is constructed when Chisel executes.
- Instead of relying on an EDSL, FIRRTL introduces a new, partially inconsistent full-custom DSL for an inter-

mediate artifact. FIRRTL therefore in turn negates some of the benefits of working with EDSLs: It is once again required to develop parsers and code generators for a new language, resulting in an increase in complexity and maintenance overhead. This paper demonstrates that the benefits that FIRRTL provides can be achieved with a model and a defined Meta-Model in the environment of an existing general-purpose programming language. It is easy to conclude that the introduction of a Meta-Model instead of the introduction of a custom language would have been beneficial.

## V. Model Driven Architecture Based Hardware Generation

The Meta-Modeling-based MDA approach presented in this paper does not exhibit these limitations and has further benefits regarding its learning curve and the domains it can be applied to. In the next paragraph, the terms modeling and Meta-Modeling are discussed. Afterwards, the concepts of MDA are introduced. Finally, a comparative analysis of MDA and HGL approaches covering the aspects introduced in Sections IV-C and IV-D is made. Focus lies hereby on the Scala-based Chisel and on Infineon's Python-based approach.

### A. Modeling and Meta-Modeling

In this paragraph, the terms modeling and Meta-Modeling are discussed and their relationship is shown. Further, possibilities of Meta-Model-based automation are described.

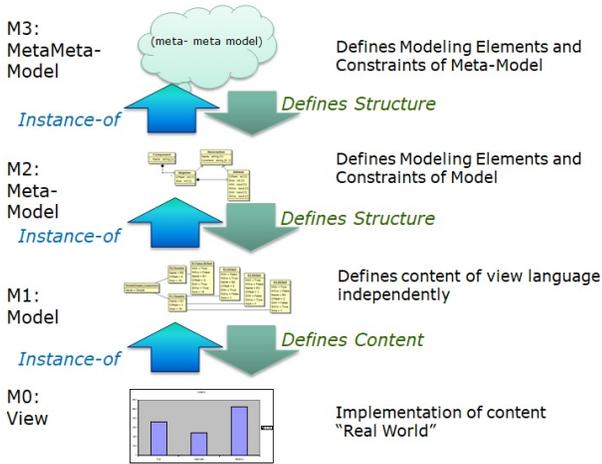

Figure 4. Meta-Modeling Stack

Figure 4 shows the standard levels of the Meta-Modeling stack. The lowest level M0 is the 'View' which describes the outputs of design collaterals of a Meta-Modeling-based generator.

The next level M1 is the 'Model'—also called data model—which is a set of structured data. The model as such is an abstract something but it may be stored and serialized in a file. The serialization of M1 does not add additional information and allows a unique reconstruction of the model. To compare with M0, an M0 model may have additional information and only makes use of parts of the M1 model. Therefore, it is in most cases not possible to uniquely reconstruct the M1 model from an M0 model.

The level above the M1 model is M2 or the Meta-Model level. It defines the structure of a model i.e. its objects, the objects' relationships and the objects' attributes as well as their types.

As the definition of a model, i.e. the Meta-Model, needs also a definition, the M3 Meta-Meta-Model is introduced. In many cases, the Meta-Model and the Meta-Meta-Model share the same formalism. This is called self-defined.

### B. Modeling and Meta-Modeling Examples

Table I shows some examples for the Meta-Modeling stack. The M0-level shows only one typical of many possible generated collaterals. The M1-M3-levels name serialized representations of the Meta*-Model.

Table I
EXAMPLES FOR META-MODELING STACK

| Level | Name | XML | UML | EMF | BNF |
|---|---|---|---|---|---|
| M3 | Meta-Meta-Model | XSD | (E)MOF | (E)CORE | EBNF |
| M2 | Meta-Model | XSD | UML | EMF | EBNF |
| M1 | Model Representation | XML | XMI | XMI | C-Grammar |
| M0 | View Example | PDF | C | Java | Parser |

To start with XML, a documentation format such as PDF is often an output of an XML-based generator stack, but as mentioned before, there may be other formats like HTML or completely documentation-independent views. M1 is serializable by XML and M2 by XSD, which is a special XML File.

Next, UML has as view example C-Code but here may be also other programming languages and coding style variants provided. A serialized UML model is stored in XMI format, which is also an XML file. Important to note is that UML supports many formalisms as activity charts or class diagrams—to name only two. These diagrams are Meta-Models and a specific activity chart or a special class diagram is a model. UML has its own formalism to describe Meta-Models called (Essential) Meta-Object-Facility (E)MOF.

The following column summarizes the underlying models of the Eclipse Modeling Framework (EMF). The view example Java has been chosen as Eclipse is implemented in this language and EMF is mainly used to make Eclipse applications. XMI is used as intermediate name. Like UML, EMF has an own formalism to define Meta-Models which is called (E)CORE.

The last column uses grammar definitions in Backus-Naur-Format (BNF) as an example. It is used to show that the concepts behind the Meta-Model stack are already quite old. As view example a parser is named and the C-grammar as one of many possible grammars. For grammar definition the Extended-Backus-Naur-Format (EBNF) is chosen as it is more convenient to write. Last but not least, as EBNF is a language on its own, (E)BNF is self-defined.

### C. Automation Starting with Meta-Models and Meta-Meta-Models

As models, Meta-Models are the basis to formally define things. In the context of the generation approach Meta-Models and Meta-Meta-Models have a much higher benefit. They can be used to automatically build wide parts of the generators' infrastructure.

Figure 5 sketches the automation that can be achieved when Meta-Models are used in the construction of generators. The lower blue part depicts the structure of a generator. A set-API is used to build a model according to a model specification. The serialization-API supports persistent storage of the model. Finally, the get-API allows a code generator, e.g. a template engine, to retrieve data from the model to make the view. This (template) engine bridges between M1 and M0 in the Meta-Modeling stack. In addition, the transformation-API offers the possibility to consistently modify a model as advanced feature, e.g. to ensure that names follow consistently a specific style.

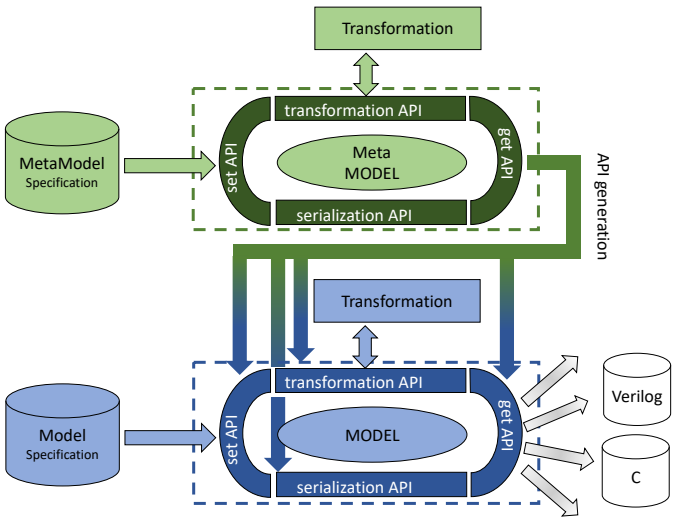

Figure 5. Automation Starting with Meta-Models

The green part is structured as the blue part but operates one level up the Meta-Modeling stack. It has the same APIs to handle the Meta-Model, i.e. it operates on M2 level. From this model, all APIs of the model-based generators can be built. Generating this APIs bridges from M2 to M1. A huge saving in coding and a guarantee for consistent coding is established in this way.

Utilizing the self-definedness of the Meta-Model and Meta-Meta-Model, the same approach can be used to build the APIs to handle the Meta-Model. Even though the savings of automatically building this API is smaller, a big new benefit arises from this approach. A set of Meta-Models and utilities can be constructed in this way starting from even more abstract specifications. To give examples, the view Meta-Models (see Figure 8) are for instance generated in this way.

### D. Embedded DSL

It might be a not neglectable effort to define the file format, parser, and (Meta-)Model builder for Meta-Models and models. An alternative way is to build an EDSL to build the model. To obtain an EDSL, an existing programming language is enriched with libraries, packages, and programming items to support an application domain in a more convenient way. As depicted in Figure 6, automatically generated APIs take a wide fraction of making the libraries etc. for the EDSL.

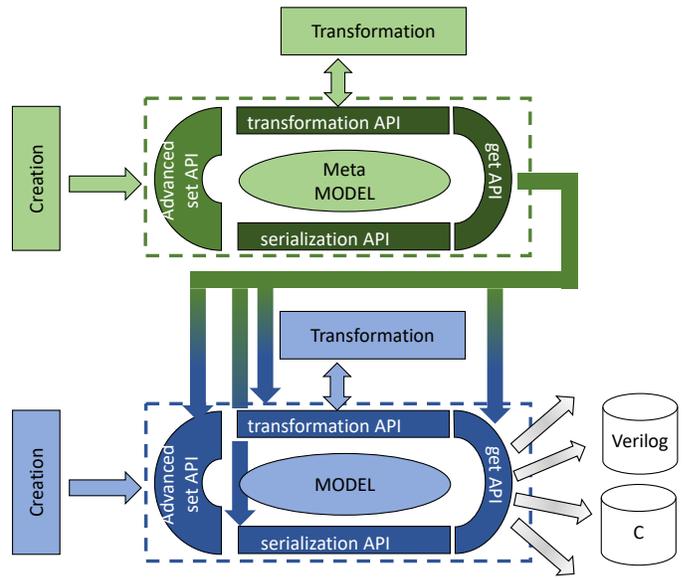

Figure 6. Utilizing Embedded DSLs

First, instead of a model and Meta-Model specification, an executable program is used to build the model/Meta-Model. To make this in a convenient way, the set-API is enhanced to allow usage of objects and and other constructs of the DSL embedding language in a more convenient way.

To give some examples: The constructor may be built similarly to instances in HDLs or overloaded assignments and operators can make a model construction look like a dataflow description.

The beauty of this approach is that the same language and language infrastructure can be used to build and to transform the model.

### E. Model Driven Architecture

The flow shown in Figures 5 and 6 in blue has to tackle a huge semantic gap between specification and implementation. As the model is mostly close to specification, the complete gap is bridged in the generator. To overcome this challenge, the Object Management Group (OMG) came up with the vision of the MDA [16].

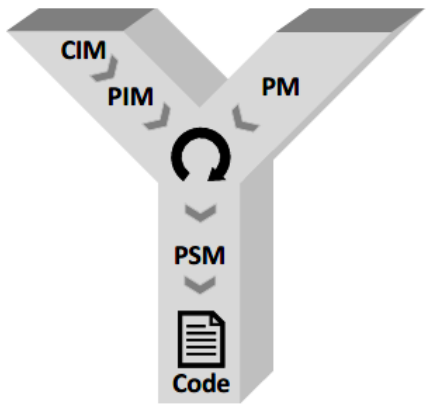

Figure 7. Y-Chart Model Driven Architecture [16]

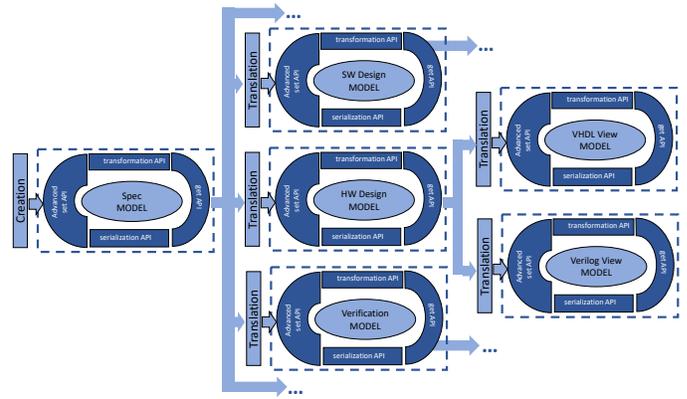

Figure 8. Model Driven Architecture Pipeline

Figure 7 depicts the general idea in a widely known MDA Y-chart. No longer one but conceptually three models are involved in the generation process. Newer versions of the MDA allow a different number than three models whereas three models are still the mostly used split. Interestingly also compilers have three models: The abstract syntax tree, the abstract machine model and the concrete machine model.

The most abstract model is called Computation Independent Model (CIM). It serves more or less the same purpose as the model in Figures 5 and 6 and represents a formalized specification.

The intermediate model is called Platform Independent Model (PIM). It represents the design independent from implementation details. The step from CIM-to-PIM consists of the definition of the targeted architecture and the mapping of algorithms on it.

The most concrete model is called Platform Specific Model (PSM). It can be constructed from a Platform Model (PM) and the Platform Independent Model (PIM).

Figure 8 depicts an implementation concept for MDA. Important to note is that this picture does not show the Meta-Models of the models even if each of the models has its own Meta-Model. Further important to note is that both, models and Meta-Models, may be composed. This increases the re-use level for both the generator construction and the generation as such and increases the consistency of translations.

A generation pipeline spans now e.g. from the specification model over the hardware design model to the Verilog view model. This last model is then unparsed to the intended view. As Figure 8 shows, the MDA environment includes a set of MDA pipelines. This is elaborated more in the next subsection.

### F. Industrial Usecases

Infineon has been using Meta-Model-based code generation for about 15 years [17]. It uses the widely adopted language Python as the underlying language. Pure XML-based approaches and preprocessors as well as template engines for VHDL have a much longer history at Infineon. Today, there is no digital design that does not make use of Meta-Model-based code generation.

The evolution to MDA started about 10 years ago [18]. To make the purpose of the models more clear Infineon called CIM Model-of-Things (MoT) reflecting the fact that things and their relationships are specified here. The 'what' is clear in focus. PIM is called Model-of-Design (MoD) as it reflects all design considerations. For digital design the MoD is called MetaRTL reflecting the chosen abstraction. PSM is called Model-of-View (MoV) as the platform dependency is introduced the way the code is generated and the libraries used by the generated code.

The MDA approach has proven its industry strength in building 30+ complex IPs and 100+ simple IPs and IP sub-modules. The result is included in many Infineon products. The approach can generate complete FPGAs or digital subsystems of SoCs but usually another IP integration framework is used for the later mentioned purpose.

As shown in Figure 8, not only hardware design but also software design and verification is automated with the MDA-approach [19]. Beyond making documentation and simulation models are also automated by following the MDA approach. Further, a set of languages can be generated as well. Besides the shown VHDL and (System)Verilog also C, C++, Java, Python and C# are supported with Rust on the roadmap. More important, depending on the kind of the language, different variants can be generated, e.g. for SystemVerilog gate-level models, RTL models and behavioral models as well as SVA properties.

### G. Analysis

The described features of an MDA-based approach are compared in this subsection with the advantages and limitations of Chisel listed in Section IV-C. As the analysis of Chisel also includes aspects of the underlying Scala language, we also include aspects of the Infineon's MDA environment underlying Python language.

*1) Disadvantage of the MDA Solution:* Even if Python has many features for generic and meta-programming, Scala is a bit better suited for making Embedded DSLs. This ends up in simpler patterns to be followed e.g. in register- and mux-inference. Since inference has some drawbacks, as discussed below, this disadvantage is more lightweight.

*2) Key Advantages of the MDA Solution:*

*a) Simplicity:* Both Chisel and the MDA approach are simpler compared to traditional, industry-standard HDLs. A big part comes from the semantic clarity of a design oriented approach over the event driven and simulation based classical HDLs. Both approaches have a concise syntax derived from proven programming languages that make the description easier.

The MDA approach is based on models which are in turn defined by their Meta-Models. The embedded domain specific language (DSL) is in wide parts generated from the Meta-Model. Hence the Meta-Model can be used as a Cheat-Sheet for both the DSL but also the transformation language and the intermediate structure (see also Section V-G2d).

Since the Meta-Model formally defines the model, a formal definition of the derived Embedded DSL exists automatically. Understanding the model and the Meta-Model is simplified in this way.

*b) Avoidance of Weakness of HDLs:* In contrast to Chisel, MDA directly maps to hardware design items. Therefore, it explicitly constructs and represents hardware related elements especially storage elements and multiplexers. No hidden inference need to be considered. Checks that guard the inference are obsolete.

Further, a model-based approach allows to offer more hardware specific constructs. So, the Infineon approach provides adders with overflow or rotate operators. Also higher level constructs widely used in hardware as state machines and lookup tables are part of the model and supported by the formalism.

*c) Easy to Learn Host Language:* Easy-to-use is essential for the hardware description formalism as its primary purpose is designing hardware.

The Infineon MDA's underlying language Python has a steep learning curve and is top in the top 20 languages [12, 13]. Python is also a complex multi-paradigm language, however, it does not require to understand all the paradigms—as functional programming—before it can be used appropriately. Further, it is interpreter-based which is said to be better suited for getting started.

Python connects to many languages, whereas the smooth integration with C/C++ is one reason why it is widely used in AI and ML. Here Python sits on top of highly efficiently implemented libraries and is mainly used to control the underlying computation. The simpleness of Python is boosted with languages that allow high efficient implementations.

*d) Visible and Clearly Defined Underlying Meta-Model:* Rephrasing the two key features of EDSL-based HGLs from Section IV-C3:

1) The ability to perform introspection on the existing data structures describing the design *during* the execution of the generator.

2) The ability to apply transformations on the generated data structures.

The MDA approach is built on models, their transformation and translation. Further, the models are immediately build when executing the model creation code. Therefore, introspection is freely available. Things like local property propagation, immediate checks, schematic viewing of an intermediate state of the design model, and generation of model parts depending on already created models are only some tasks that can be done while building the model.

*e) Serialization:* The model defined by the Meta-Model is the internal model but also the external representation. Auto-generated (de-)serializes guarantee that the external representation contains the same information as the internal model.

*f) Bridging the Semantic Gap:* MDA works with three models and thus splits bridging the semantic gap by separation of concerns. Further, reusable transformers can be build on every kind of model to apply design level specific but general applicable adaptations. To name two examples: Transformations on the design level can harden flip-flops to decrease the error rate of designs. Transformations on the language level can adopt the generated collateral to specific coding styles. In addition, formatting and comment insertion is done in the final unparse step, thus separating this annoying yet important step.

All intermediates follow the same concepts and support the same operators, types and expressions in the same way. As the complete infrastructure is auto-generated, the overhead of defining three and more Meta-Models to handle the models is kept at an acceptable level.

*g) Support of Multi-Domains, Multi-Purpose and Multi-Languages:* As mentioned above, MDA cannot only be used for digital design creation. Through its general concept, it can—and is at Infineon—applied to generation in the software and mixed hardware/software domain as well. It is not only applied to design code, but also to automate verification, building of abstract models, and documentation.

In order to keep the overall effort low, to support cross-domain consistency and to enable a smooth integration of collaterals of different domains, Infineon uses specific type- and expression-Meta-Models that are used as sub-Meta-Models in all the domains mentioned above.

## VI. Summary and Outline

This paper shows that an MDA-based approach can provide extremely powerful generators with reasonable implementation effort. The paper provides an in-depth comparison of the Meta-Modeling approach with HGLs, especially Chisel. The comparison highlights unique advantages of a Meta-Modeling-based approach and summarizes the benefits. Finally, it concludes that this approach is a superior alternative to the most advanced HGLs such as SpinalHDL and Chisel.

Future potential lies in widening the approach in direction of e.g. requirement engineering, analog- and analog-mixed-signal design, or design for test.

## VII. ACKNOWLEDGMENTS

This work summarizes the EDCC'2024 keynote talk entitled *4th Generation HW-Design-Formalisms* (https://edcc2024.esat.kuleuven.be/keynotes.html). A big thank you to the EDCC organizers to give the chance to present the topic.

The work had been partially supported by the German Federal Ministry of Education and Research (BMBF) and ITEA within the project GenerIoT under contract number 01IS22084A.

Sections II, III, and IV of this paper have been previously published in a similar form in [20].